\documentclass[reprint,aps,pra,twocolumn,showpacs,superscriptaddress,preprintnumbers,amsmath,amssymb]{revtex4-1}
\usepackage{bm}
\usepackage{amsmath}
\usepackage{amssymb}
\usepackage{graphicx}
\usepackage{epsfig}
\usepackage{color}
\def\be{\begin{equation}}
\def\ee{\end{equation}}
\def\bea{\begin{eqnarray}}
\def\eea{\end{eqnarray}}
\def\f{\frac}

\def\nn{\nonumber}
\def\no{\noindent}
\def\ra{\rangle}
\def\la{\langle}

\def\tU{{\widetilde{U}}}

\begin{document}

\title{Detection of a quantum particle on a lattice 
under repeated projective measurements }

\author{Shrabanti Dhar}
\email{shrabantidhar@gmail.com}
\author{Subinay Dasgupta}
\affiliation{Department of Physics, University of Calcutta, 92 Acharya Prafulla Chandra Road, 
Kolkata 700009, India}
\author{Abhishek Dhar}
\affiliation{International Centre for Theoretical Sciences, TIFR, 
Bengaluru 560012, India}
\author{Diptiman Sen}
\affiliation{Centre for High Energy Physics, Indian Institute of Science, 
Bengaluru 560012, India
}
\date{\today}

\begin{abstract}
We consider a quantum particle, moving on a lattice with a tight-binding
Hamiltonian, which is subjected to measurements to detect it's arrival
at a particular chosen set of sites. The projective measurements are 
made at regular time intervals $\tau$, and we consider the evolution of 
the wave function till the time a detection occurs. 
We study the probabilities of its first detection at some time and conversely 
the probability of it not being detected (i.e., surviving) up to that time. 
We propose a general perturbative approach for understanding the dynamics 
which maps the evolution operator, consisting of unitary transformations 
followed by projections, to one described by a non-Hermitian Hamiltonian. For 
some examples, of a particle moving on one and two-dimensional lattices with 
one or more detection sites, we use this approach to find exact expressions 
for the survival probability and find excellent agreement with direct 
numerical results. A mean field model
with hopping between all pairs of sites and detection at one site is solved 
exactly. For the one- and two-dimensional systems, the survival probability 
is shown to have a power-law decay with time, where the power depends on 
the initial position of the particle. Finally, we show an interesting and 
non-trivial connection between the dynamics of the particle in our model 
and the evolution of a particle under a non-Hermitian Hamiltonian with 
a large absorbing potential at some sites. 
\end{abstract}

\maketitle 

\section{Introduction}

The measurement of the arrival time of a quantum mechanical particle in a 
given detection region is a longstanding and fundamental problem in quantum 
mechanics \cite{allcock69,kijowski74,kumar85,grott96,aharonov98,savvidou06,mugap00,damborenea02,galapon04,galapon05,muga08,
yearsley,savvidou12,vona13,muga00,vega1,vega2}. In spite of much effort, the construction of a time 
operator has been found to be controversial \cite{mielnik}. 
One possible approach that one could take to find the time of 
arrival is the following. Suppose that a particle is released from a given 
region at time $t=0$ and is allowed to evolve unitarily with some Hamiltonian. 
In some specified detection region we make repeated instantaneous, projective 
measurements, at regular time intervals $\tau$, to see if the particle has 
arrived there, and we stop once a detection is made. If the detection occurs 
at the $n^{\rm th}$ measurement, we could say that the particle's time of 
arrival into the detection region, is at time $t=n \tau$ (or more precisely 
between times $t-\tau$ and $t$). Repeated measurements of this kind are known 
to have a somewhat surprising feature when one considers the limit where the 
time interval between measurements $\tau$ is taken to zero. One finds that 
the probability of detecting the particle goes to zero; this is called the 
quantum Zeno effect \cite{misra77,shimizub05,facchi08,wineland90,kofman,kwiat,cirac,itano09,signoles}.

An interesting case to consider is one where the time $\tau$ between 
measurements is 
assumed to be small compared to the typical spreading time of a wave packet 
(in the models that we will study, the spreading time is of the order of 
$1/\gamma$, where $\gamma$ is the hopping amplitude between nearest neighbors),
but {\emph {is kept finite and we do not take the limit $\tau \to 0$}}. 
The problem of the effect of repeated measurements, made at finite 
time intervals, on the evolution of a quantum system has been studied in 
various contexts both theoretically and experimentally 
\cite{hegerfeldt96,erez08,jahnke,christian12,halliwell10,ingold11}.
Because of the probabilistic nature of the quantum detection process, we 
expect the time of detection to be a stochastic variable. The probability 
distribution of this time of first detection of the particle in some given 
region, and the complementary probability of not being detected (i.e., 
survival) at all up to some time, are then interesting quantities to study. 
The time evolution of the wave function of a surviving particle is also of 
interest. 
 
In an earlier paper \cite{dhar13}, we addressed this problem taking the 
example of a particle moving on a one-dimensional lattice with a tight-binding 
Hamiltonian, with detections made at a single site. This paper extends our 
earlier work in several directions. We consider here 
the motion of a single particle on an arbitrary lattice, with the dynamics 
still controlled by a tight-binding Hamiltonian. Also, we allow our projective 
measurements to be made on more than one site in a region. Through a general 
perturbative treatment, valid when the time interval between measurements is 
small compared to the wave spreading time scale, we show that the long time 
dynamics of the particle can be effectively described by a non-Hermitian 
Hamiltonian. The non-Hermitian Hamiltonian is defined on the subspace 
consisting of non-measurement sites and contains a small imaginary potential 
on all the sites that are connected directly by hopping to the measurement 
sites. Using this result, we are able to solve for the time evolution of 
initially localized wave functions and from this, analytically compute the 
survival probability for several examples of particles in one- and 
two-dimensional lattices. Our analytic results are compared with 
direct numerical results and we find excellent 
agreement. We find that the survival probability decays as a power of 
the time, where the power depends on the initial position of the particle. 
Finally, we demonstrate another mapping that can be made between 
the effective non-Hermitian Hamiltonian with a 
small imaginary potential that appears in our model, and a different 
non-Hermitian Hamiltonian which has a large imaginary potential on the 
measurement sites. This second non-Hermitian Hamiltonian is similar to what 
has been proposed in the context of the study of the time of arrival of a 
free quantum particle (moving in continuous space) into a given region 
(the half line), using the approach of repeated measurements 
\cite{muga08,halliwell10}. This also relates our study to a recent work of 
Krapivsky {\emph {et al}} ~\cite{mallick13} who look at the survival 
probability of a particle moving on a one-dimensional lattice with 
imaginary potentials at one or more sites.

The plan of the paper is as follows. In Sec.~\ref{sec:model}, we describe 
our precise model and the repeated measurement protocol. We show that the 
measurement dynamics is described by an effective non-unitary evolution 
operator which evolves the wave function between successive measurements. 
We explain how an expression for the survival probability $P_n$ after 
the $n^{\rm th}$ measurement can be obtained from the wave function.
In Sec.~\ref{sec:pert} we describe the perturbation theory by which we are 
able to describe the non-unitary dynamics by an effective non-Hermitian 
Hamiltonian. In Sec.~\ref{sec:examples} we consider several examples of 
a single quantum particle moving on one- and two-dimensional lattices and 
described by tight-binding Hamiltonians with nearest neighbor hopping terms,
which is subjected to regular measurements made on one or more sites at 
regular intervals of time $\tau$. We derive perturbative results 
for the survival probability and the effective wave function of the particle 
after a time given by an integer multiple of $\tau$. Analytical and numerical 
results are presented. We also present (Sec.~\ref{sec:mf}) an exact solution 
for a mean field type of model where the particle can hop from any site to all 
the other sites. Sec.~\ref{sec:pert2} describes the mapping between the 
non-Hermitian Hamiltonian problem with a large imaginary potential at the 
measurement sites and our problem with small measurement time intervals.
We conclude with a discussion in Sec.~\ref{sec:summary}.

\section{Model and general framework}
\label{sec:model}
Our model consists of a particle moving on a discrete lattice of $N$ sites 
and its dynamics is described by a tight-binding type Hamiltonian of the form
\bea H= \sum_{\substack{r,s =1 }}^N H_{r,s}~|r\ra \la s|~, \label{ham} \eea
where $H$ is taken to be a real symmetric matrix whose non-vanishing elements 
have strength $\gamma$ (in units of $\hbar$). The free time evolution of 
$|\psi\ra $ is given by 
\be |\psi (t) \ra = U_t |\psi (0)\ra,~~{\rm where}~~ U_t=e^{-i H t}~. \ee
Let us define the projection operator $ {A}= \sum_{r \in D} |r \ra \la r|$ 
corresponding to a measurement to detect the particle in the domain $D$ 
containing a fixed number $N_D$ of sites, and the complementary operator $B=
1-A$ corresponding to the projection to the space of $N_S$ sites belonging to 
the ``system''. According to the measurement postulate of quantum mechanics, 
the probability of detecting the particle on performing a measurement on the 
state $|\psi \ra$ is $ p = \sum_{r \in D} | \la r | \psi \ra |^2 = \la \psi | 
A | \psi \ra$. The probability of non-detection or the {\emph{ survival 
probability}} is then $P= \la \psi | B | \psi \ra =1-p$. The measurement 
postulate also tells us about the state of the system immediately after the 
measurement. They in fact alter the Hamiltonian time evolution of the 
system. If the measurement detects the particle (with probability 
$p$) then the state after measurement is $A|\psi\ra$, while if a measurement 
{\emph{does not}} detect the particle (with probability $P$), then the 
state immediately after measurement is $|\psi^+\ra= B|\psi\ra$, with 
appropriate normalizations. Thus we see that after the measurement the system 
is effectively described by a density matrix $\rho'=A |\psi\ra \la \psi 
|A + B |\psi\ra \la \psi |B$. However in our scheme {\emph{we stop the 
experiment whenever a particle is detected}}. Hence only those states that 
are projected onto the system subspace are further evolved, and we do not 
need to consider density matrices. After the first measurement we again 
unitarily evolve the state $|\psi^+\ra= B|\psi\ra$ until the next measurement.

We consider a sequence of measurements $n=1,2\ldots$ at intervals of time 
$\tau$ which continue until a particle is detected. \emph{Thus the time 
evolution is given by a sequence of unitary evolutions followed by projections,
onto the subspace corresponding to $B$, till the particle is detected}.
Let $|\psi^-_n\ra$ and $|\psi^+_n\ra$ be the (un-normalized) wave functions
of the system, immediately before and after the $n^{\rm th}$ measurement 
respectively. We note that $|\psi^-_n\ra = U_\tau|\psi^+_{n-1}\ra$ and 
$|\psi^+_n\ra=B|\psi^-_n\ra$. Hence, defining $\tU=BU_\tau$, it follows that
\bea |\psi^-_n \ra =U_\tau \tU^{n-1} |\psi (0) \ra ~~~{\rm and}~~~
|\psi^+_n \ra = \tU^{n} |\psi (0) \ra~. \label{evolpsi} \eea
Let $P_n$ be the probability of survival after $n$ measurements. Then clearly
\bea P_1 = \la \psi^-_1|B | \psi^-_1 \ra= \la \psi(0)| \tU^\dagger \tU |
\psi(0) \ra=\la \psi^+_1|\psi^+_1\ra~. \nn \eea
Note that $P_1$ is thus the normalizing factor for $|\psi^+_1\ra$ and also 
for $|\psi^-_2\ra$. The survival probability after the second measurement is 
obtained as the product of non-detection at $n=1$ times the probability of 
non-detection at $n=2$, and is given by
\bea P_2 =P_1 \times \f{ \la\psi^-_2 |}{\sqrt{P_1}} B \f{|\psi^-_2 \ra}{
\sqrt{P_1}} =\la \psi(0)| \tU^{\dagger 2} \tU^2 |\psi(0) \ra=\la \psi^+_2|
\psi^+_2\ra~.\nn \eea
Proceeding iteratively in this way, we get 
\bea P_n =\la \psi(0)| \tU^{\dagger n} \tU^n |\psi(0) \ra=\la \psi^+_n|
\psi^+_n\ra~.\label{PS} \eea
If we imagine an ensemble of identically prepared states on which we perform 
repeated measurements, then $P_n=\la \psi^+_n|\psi^+_n \ra$ gives the fraction 
of systems for which there has been no detection and that are still evolving.
The probability of first detection in the $n^{\rm th}$ measurement is given by
\begin{align}
p_n&=P_{n-1} \times \f{ \la\psi^-_n |}{\sqrt{P_{n-1}}} A \f{|\psi^-_{n} \ra}{
\sqrt{P_{n-1}}}
=\la \psi^-_n|A|\psi^-_{n}\ra~ \label{pdet} \\
&=\la \psi^-_n|\psi^-_{n}\ra- \la \psi^-_n|B|\psi^-_{n}\ra=P_{n-1}-P_n~,
\end{align}
as expected.

Our main aim in the rest of the paper is to study the behavior of the 
survival probability for different cases. From the discussion above it 
is clear that the central problem is to understand the properties of the 
effective evolution operator $\tU$ which, for initial states located inside 
the ``system'', is equivalently given by 
\be \tU \equiv B e^{-i H \tau} B~. \ee
An explicit diagonalization of this non-Hermitian evolution operator is 
difficult in general. In the next section we provide a perturbative approach, 
valid for small $\tau << 1/\gamma$. As we will see from our numerical
results, this gives a quite accurate description.
 
\section{Perturbation theory and connection to an effective 
non-Hermitian Hamiltonian}
\label{sec:pert}
Let us use the following notation: we divide our full set of sites into those 
belonging to the ``system'' (labeled by roman indices $l,m$) and those 
belonging to the domain $D$ consisting of sites where measurements are made 
(denoted by greek indices $\alpha, \beta$). With this notation we have $A=
\sum_{\alpha} |\alpha\ra \la \alpha |$ and $B=\sum_{l} |l \ra 
\la l |$, while the Hamiltonian in Eq.~(\ref{ham}) can be rewritten in the 
following form
\begin{align}
H&=H_S+H_M+V~, \label{ham2} \\
{\rm where}~~H_S&=\sum_{l,m} H_{l,m} |l\ra \la m |,~~~ H_M= \sum_{\alpha,\beta} 
H_{\alpha,\beta} |\alpha \ra \la \beta|, ~~~\nonumber\\
 {\rm and}~~V&= \sum_{l,\alpha} 
V_{l,\alpha} |l\ra \la \alpha | + V_{\alpha,l} |\alpha \ra \la l |, \nn
\end{align}
describe the system, measurement sites and coupling parts respectively of the 
full Hamiltonian.

Expanding the effective evolution operator $\tU = B e^{-iH t} B$ to second 
order in $\tau$ gives in the system subspace,
\begin{align}
\tU&= B~\left[~I-iH \tau -\f{\tau^2}{2} H^2+\ldots~\right]~B \nn \\
&=I-iH_S \tau -\f{\tau^2}{2} H_S^2 \nonumber \\
&-\f{\tau^2}{2} \sum_{l,m} \sum_\alpha 
V_{l,\alpha}V_{\alpha,m} |l\ra \la m| + \ldots \nn \\
&=e^{-i H_{eff} \tau}+{\mathcal{O}}(\tau^3), \nn \\
{\rm where}~~H_{eff}&=H_S+V_{eff}, \label{effH} \\
{\rm and}~~V_{eff}&=-\f{i \tau}{2} \sum_{l,m} \sum_\alpha V_{l,\alpha}
V_{\alpha,m} |l\ra \la m| ~. \nn
\end{align}
($I$ denotes the $N_S \times N_S$ unit matrix).
Thus we see that our system is effectively described by a non-Hermitian 
Hamiltonian $H_{eff}$ and the problem now reduces to diagonalizing this 
Hamiltonian. Note in particular that the strength of the non-Hermitian 
potential is small and proportional to the measurement interval $\tau$. 
In the following section we will give explicit examples on 
regular lattices in one and two dimensions where this reduced problem can be 
tackled analytically, and comparisons can be made with direct numerical 
solutions of the original problem. 
\vspace{0.05 cm}
\section{Comparisons between perturbation theory and direct 
numerical results}
\label{sec:examples}
\subsection{Particle in a one-dimensional box}

We consider the motion of a quantum particle in a one-dimensional 
lattice with only nearest neighbor hoppings and open boundary conditions. 
The corresponding full Hamiltonian is
\[ H = - ~\gamma ~\sum_{l=1}^{N-1} \left( ~|l+1 \ra \la l | + |l
\ra \la l +1 | ~\right)~. \]
Without loss of generality we set $\gamma=1$. We consider three different cases
corresponding to different choices of the measurement points: (i) $\alpha = 
N$, (ii) $\alpha =N-N_D+1,N-N_D+2,\ldots,N$ and (iii) $ \alpha=1,N$.\\

\no {\emph{Case (i): Measurement at single point at one end of the box.}}\\
This case was presented elsewhere \cite{dhar13} but we present it here again 
as an illustrative example of our present general framework. 
The projection operators now are ${A}=|N\ra \la N|$ and $B= \sum_{l=1}^{N-1} 
|l\ra \la l |$. From the general discussion in Sec.~\ref{sec:pert} it is 
clear that the effective Hamiltonian for the $N-1$ sites system is given by
\begin{align}
H_{eff} &= H_S+V_{eff}, \nn \\
{\rm where}~~H_S =& - \sum_{l=1}^{N-2} \left( ~|l+1 \ra \la l | + |l 
\ra \la l +1 | ~\right)~~{\rm and}\nonumber\\
&V_{eff}=-\f{i\tau}{2}|N-1\ra \la N-1|~.
\label{H_add}
\end{align}
We now obtain the eigenvalues and eigenvectors of this effective Hamiltonian 
using first order perturbation theory. The eigenvalues of $H_S$ (with $N-1$ 
sites) are given by
\be \epsilon_s=-2\cos \left(\f{s\pi}{N}\right), \label{eigenH1} \ee
while the eigenvectors $\la l|\phi_s \ra = \phi_s(l)$ are given by
\be \phi_s(l)=\sqrt{\f{2}{N}} \sin\left( \f{s l \pi}{ N} \right)
\label{eigenH2} \ee 
for $s=1,2,\ldots,N-1$. Treating $V_{eff}$ as a perturbation, we find the 
following modified spectrum for $H_{eff}$,
\begin{align}
\mu_s &=\epsilon_s+\la \phi_s| V_{eff} |\phi_s\ra=\epsilon_s-\f{i}{2} 
\alpha_s, \\
{\rm with}~~\alpha_s &= \f{2 \tau}{N} \sin^2\left(\f{s \pi}{N}\right)~.
\end{align} 
This means that an eigenstate of $H_S$ will decay with time and, after $n$ 
measurements made at times $t=n \tau$, the state of the system is given by
\[ |\phi_s(t)\ra = e^{-i H_{eff} t} |\phi_s \ra= e^{-\alpha_s t/2} e^{-i 
\epsilon_s t} |\phi_s\ra. \] Hence the survival probability $P_s(t)$ is 
given by the exponential decay 
\[ P_s(t) = \la \phi_s(t)|\phi_s(t)\ra = e^{-\alpha_s t}~, \]
and the first detection probability by $p(t)=-dP/dt = \alpha_s e^{-\alpha_s 
t}$. Since the decay rate $\alpha_s$ depends on $\tau$, it vanishes in the 
limit $\tau \to 0$, and one obtains the {\emph{quantum Zeno effect}}.

\begin{figure}
\centering
\hspace{-0.1cm}\includegraphics[clip,width=3.48in, angle=0]{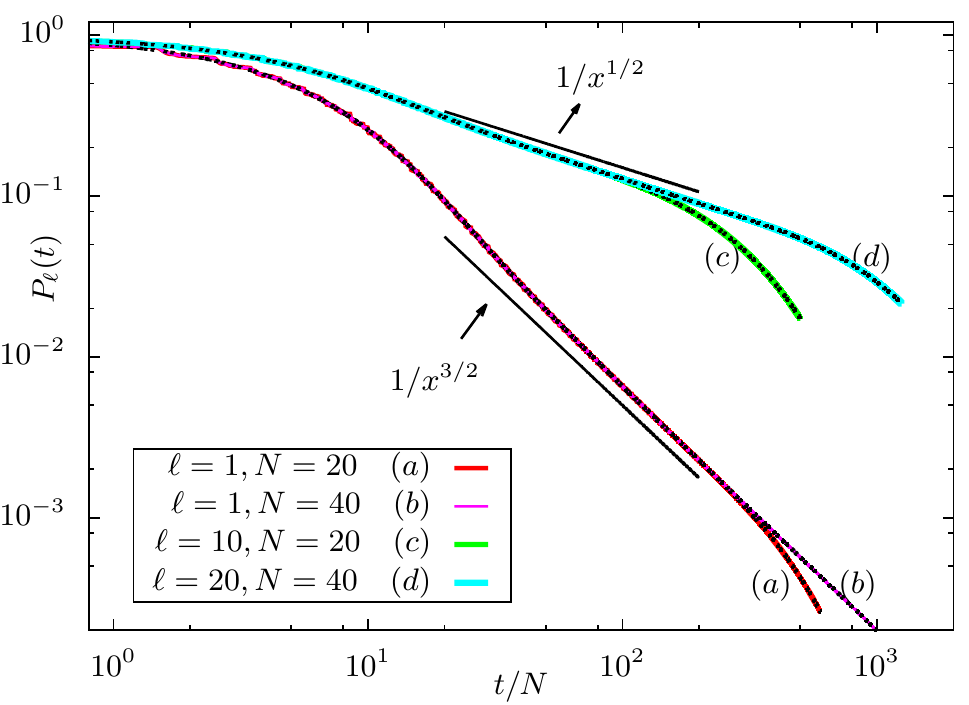}
\caption{(Color online) Open Boundary Conditions: Decay of the survival probability 
$P_\ell(t)$ for different initial position egenstates. The black dotted 
lines are the predictions from perturbation theory. The solid black lines are 
the predicted power law decays for initial points in the bulk and at the 
boundary. The measurement was done at site $N$, and the measurement time 
interval was taken to be $\tau=0.1$.}
\label{obc}
\end{figure}

\begin{figure}
\hspace{1.5cm}{\includegraphics[clip,width=9cm, angle=0]{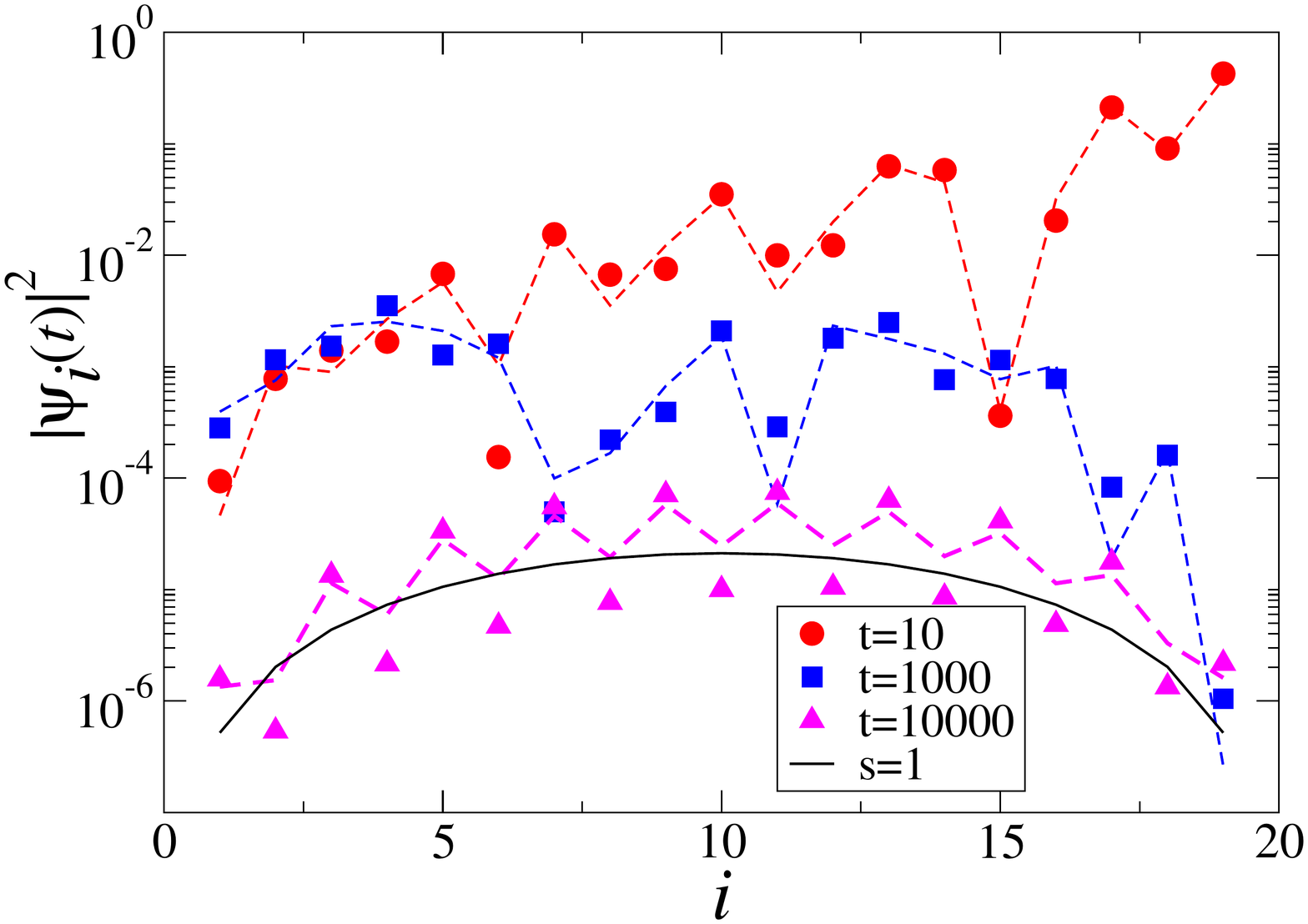}}
{\includegraphics[clip,width=9cm, angle=0]{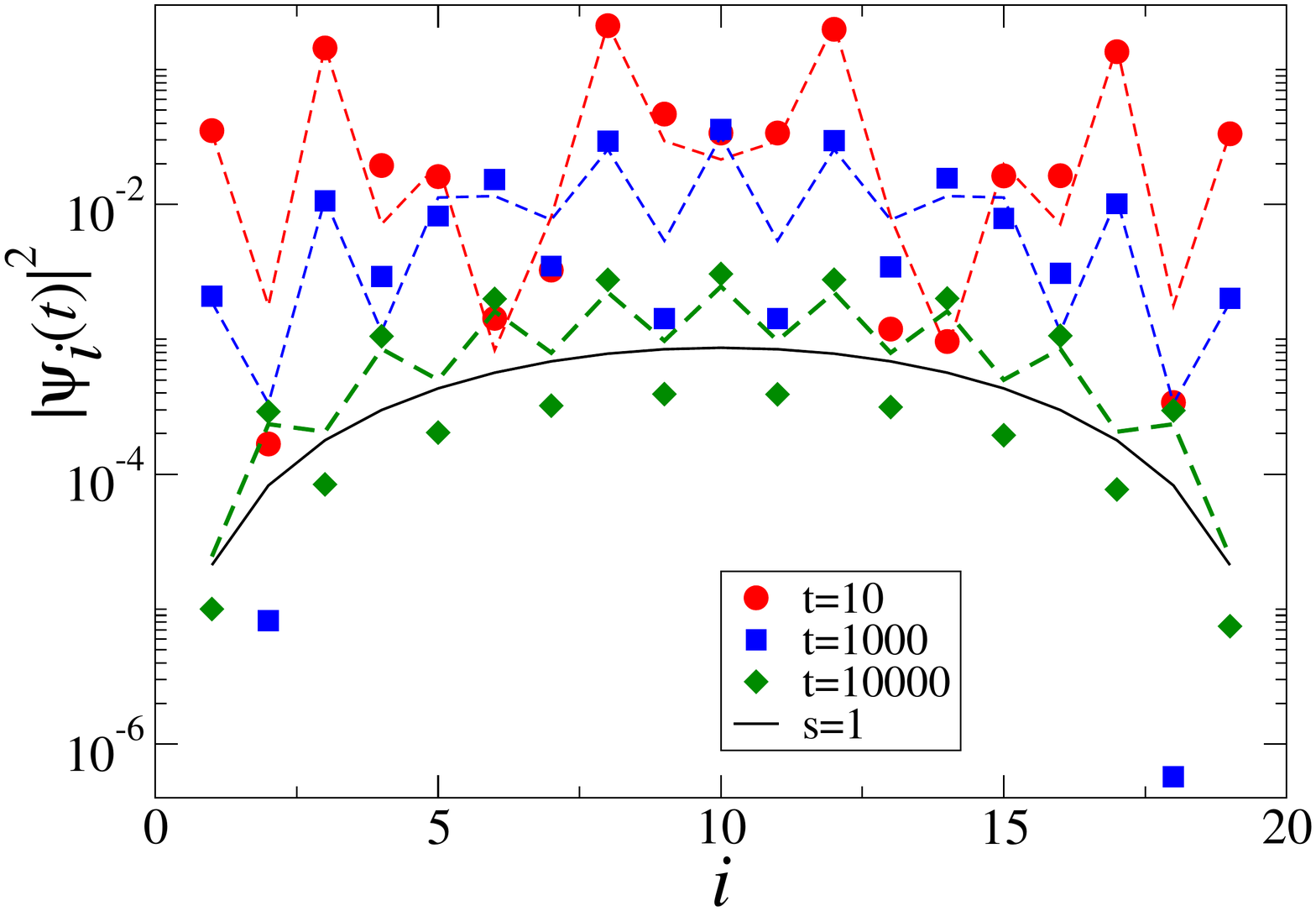}}
\caption{(Color online) Open Boundary Conditions: Plot of the probability density 
$|\psi_i(t)|^2$ at different times when the initial state is a position 
eigenstate with (top)$\ell=1$ and (bottom)$\ell=10$. The 
dashed lines are the predictions from the perturbation theory 
[Eq.~(\ref{wfob})]. The solid line is the plot for the lowest eigenstate. 
The other parameters were $N=20$ and $\tau=0.1$.}
\label{psiop_l}
\end{figure}

When the initial state is a position eigenstate , it has been shown 
\cite{dhar13} that 
the time evolution is given by 
\be | \psi(t) \ra = e^{-i H_{eff} t} |\ell\ra= \sum_s \phi_s(\ell) 
e^{-\alpha_s t/2} e^{-i \epsilon_s t} |\phi_s \ra~, \label{wfob} \ee
and thus the survival probability becomes
\begin{align}
P_\ell(t)&=\la \psi(t) | \psi(t)\ra = \sum_{s=1}^N \f{2}{N} \sin^2 (
\f{s \pi \ell}{N}) e^{-\f{2 \tau t}{N} \sin^2 (\f{s \pi}{N})}~. \label{Pit} 
\end{align}
For large $N$ and in the time window where $t\tau/N$ is large but $t\tau/N^3$ 
is small, Eq.~ (\ref{Pit}) becomes
\be P_{\ell} (t)= \f{1}{\sqrt{8\pi x}}\left[1-e^{-\ell^2/2x}\right], ~~~{\rm 
where}~~~ x = \f{t\tau}{N}. \label{Pcont1} \ee
If the particle was initially close to the left boundary ($\ell \sim O(1)$), 
then the survival probability decays as $1/t^{1/2}$ for small $t$ and as 
$1/t^{3/2}$ for large $t$. On the other hand, if the particle was initially 
well within the bulk ($\ell \sim O(N)$), then one observes only the former 
behavior of $1/t^{1/2}$. At times $t \sim N^3$ there is an exponential decay 
with time. We show in Fig.~\ref{obc} the decay of the survival probability 
with time, as computed numerically from the exact expression in Eq.~ (\ref{PS})
and from the analytical perturbative expression in Eq.~(\ref{Pit}).
It is clear that there is very good agreement between the direct 
results and those obtained from perturbation theory. The form of the 
wave functions at different times is shown in Fig.~\ref{psiop_l}. We see 
that at large times ($O(N^3)$) the wave function gets a contribution mainly 
from the lowest eigenstate and one can understand the exponential decay at 
these time scales. Another interesting feature is that the behaviors for 
$\ell$ and $N-\ell$ are the same, $P_{\ell}(t) =P_{N-\ell}(t)$, due to the 
symmetry $\phi_s({\ell})=\phi_s({N-\ell})$.\\

\no \emph{Case (ii): One-dimensional box with measurements done on 
several sites at one end.}

In this case the measurement projection operator is given by $A=
\sum_{\alpha=N-N_D+1}^N |\alpha \ra\la \alpha|$ and the system consists of 
$N_S=N-N_D$ points. We notice that, because of the nearest neighbor coupling 
form of the Hamiltonian, the form of $V_{eff}$ is now given by $V_{eff}=-i 
(\tau/2) |N-N_D><N-N_D|$. Hence the analysis of the previous case remains 
valid with the simple replacement $N-1 \to N-N_D$. In particular,
we recover the same asymptotic behavior for the survival probability. 
Physically we can understand this result as follows --- since $\tau$ is small, 
the particle can propagate only up to one site during time $\tau$. Thus the 
systems with $1$ and $N_D$ measurement sites at the end are the same (for 
$N_D>1$) as the particle never visits sites beyond the first detector site.\\ 

\no \emph{Case (iii): One-dimensional box with measurement done at boundary 
sites on both ends.}
 
The measurement projection operator now is $A=|1\ra\la 1|+|N\ra \la N|$ and the
number of sites on the system is $N-2$. The effective interaction is then 
given by $V_{eff}=-(i \tau/2)( |2\ra\la 2|+|N-1\ra \la N-1|)$. The eigenstates 
of $H_S$ (with $N-2$ sites) are now given by 
\[ \phi'_{s}(l)=\sqrt{\f{2}{N-1}}\sin\left(\f{s \pi (l-1)}{N-1}\right),\]
with $s=1,2,\ldots,N-2$ and $l=2,3,\ldots,N-1$, and the eigenvalues of 
$H_{eff}$ are given by $\mu_s=\epsilon'_s -(i/2) \alpha'_s$, with $\epsilon'_s
=-2\cos[s \pi/(N-1)]$ and $\alpha'_s =\tau [\phi'^{2}_s(2)+ \phi'^{2}_s(N-1)]=
[4\tau/(N-1)]~\sin^2 [s \pi/(N-1)]$. Thus, for large $N$, we get a decay 
constant for the eigenstates which has twice the value of that in case~(i), 
corresponding to the fact that there is absorption at two boundary points. 
Clearly the asymptotic results given for case~(i) for survival probability of 
initial position eigenstates continue to hold. 

\subsection{One-dimensional lattice with periodic boundary conditions}

We now consider a ring geometry with a Hamiltonian given by
\be {H} = -\sum_{l=1}^{N} \left(~ |l+1 \ra \la l | + |l \ra 
\la l+1 |~ \right)~ \ee
with $|N+1 \ra \equiv |1\ra$, and we assume that $N$ is even. This case was 
also presented in Ref. \cite{dhar13} and we discuss it now within the present
framework. Taking $A= |N\ra\la N|$ we get $H_S= \sum_{l=1}^{N-2} ~(|l+1 \ra 
\la l | + |l \ra \la l+1 |)$ (the same as that for Case (i)), 
and $V_{eff}=(-i \tau/2) (|N-1\ra \la N-1|+|1\ra \la 1|+|N-1\ra 
\la 1|+|1\ra \la N-1|)$. 

This case is somewhat special because it turns out that half of the energy 
eigenstates of $H$ are unaffected by the measurement process. To see this we 
notice that there are $(N-2)/2$ eigenvalues of $H$, given by 
$e_s=-2\cos (2s\pi/N)$, for $s=1,2,\ldots,N/2-1$, which are two-fold 
degenerate with eigenvectors 
\be \psi_s(l)=\sqrt{\f{2}{N}} \sin\f{2s l \pi }{N} ~~~~{\rm and}~~~~ 
\psi_{s+N/2 - 1}(l)= \sqrt{\f{2}{N}} \cos\f{2s l \pi }{N}. 
\label{eigenH3} \ee
The remaining two eigenvalues $-2$ and $2$ correspond respectively to 
eigenvectors 
\[ \psi_{N-1}(l)=(-1)^{l}/\sqrt{N} ~~~~{\rm and}~~~~ \psi_N(l) = 1/\sqrt{N}. \]
We now observe that $\psi_s(l)$, for $s=1,\ldots,N/2-1$, are also exact 
eigenstates of the Hamiltonian for an open chain with $N-1$ sites $l=1,\ldots,
N-1$. Let us denote these eigenstates of $H_S$ by $\phi_{2s}(l)$ and the 
corresponding eigenvalues by $\epsilon_{2s}=e_s$. Since $\phi_{2s}(l)$ vanish 
at the detector site $l=N$, they are not affected by the measurements and 
are exact eigenfunctions of $\widetilde{U}$. The rest of the eigenstates of 
$H_S$, given by $\phi_{2s+1}(l)=\sqrt{{2}/{N}} \sin [(2s+1)l\pi/{N}]$ with 
eigenvalues $\epsilon_{2s+1}=2 \cos [(2s+1)\pi/N] $, for $s=0,1,\ldots,N/2-1$, 
decay with a rate which we can again compute from perturbation theory. We find 
\be \alpha_{2s+1} = - \frac{2}{i} \la\phi_{2s+1}|V_{eff}|\phi_{2s+1}\ra = 
4 \tau \phi^2_{2s+1} (1). \label{defQZ2} \ee 

If the particle is initially at site $\ell$, its time evolution 
is given by
\begin{align} 
| \psi(t) \ra &= \sum_{s=1}^{N/2-1} \phi_{2s}(\ell) e^{-i \epsilon_{2s} t} |
\phi_{2s} \ra ~+\nonumber \\
& \sum_{s=0}^{N/2-1} \phi_{2s+1}(\ell) P^{1/2}_{2s+1}(t) 
e^{-i \epsilon_{2s+1} t} |\phi_{2s+1} \ra, \label{wfpb} 
\end{align} so that,
\be P_\ell(t)-P_\ell(\infty) =\sum_{s=0}^{N/2-1} \phi_{2s+1}^2(\ell) 
P_{2s+1}(t). \label{Padd1}\ee
For large $N$ and large $t\tau/N(=x)$, Eq. (\ref{Padd1}) becomes
\begin{align} 
P_\ell(t)-P_{\ell}({\infty})&= \f{1}{2 \pi} \int_{-\infty}^{\infty} dq 
\sin^2 (q \ell) e^{- 8xq^2} \nonumber \\
&=\f{1}{ 8 \sqrt{ 2\pi x}}\left[1-e^{-\ell^2/ 8x}
\right], \label{Pitcont2}
\end{align} 
with $P_l(\infty) = \sum_{s=1}^{N/2-1}\phi_{2s}^{2}(\ell)$ which is equal to
$1/2$ for $\ell \neq N/2$ and zero for $\ell=N/2$. Thus, $P_\ell(t)-
P_{\ell}(\infty)$ decays as $1/t^{1/2}$ when the initial position $\ell$ is in 
the bulk, and as $1/t^{3/2}$ when $\ell$ is near the boundary. Furthermore, 
when $\ell=N$ the decay is of order $1/t^{1/2}$ and $P_N(\infty)=0$.

\begin{figure}
\hspace{-0.1cm}\includegraphics[clip,width=3.45in, angle=0]{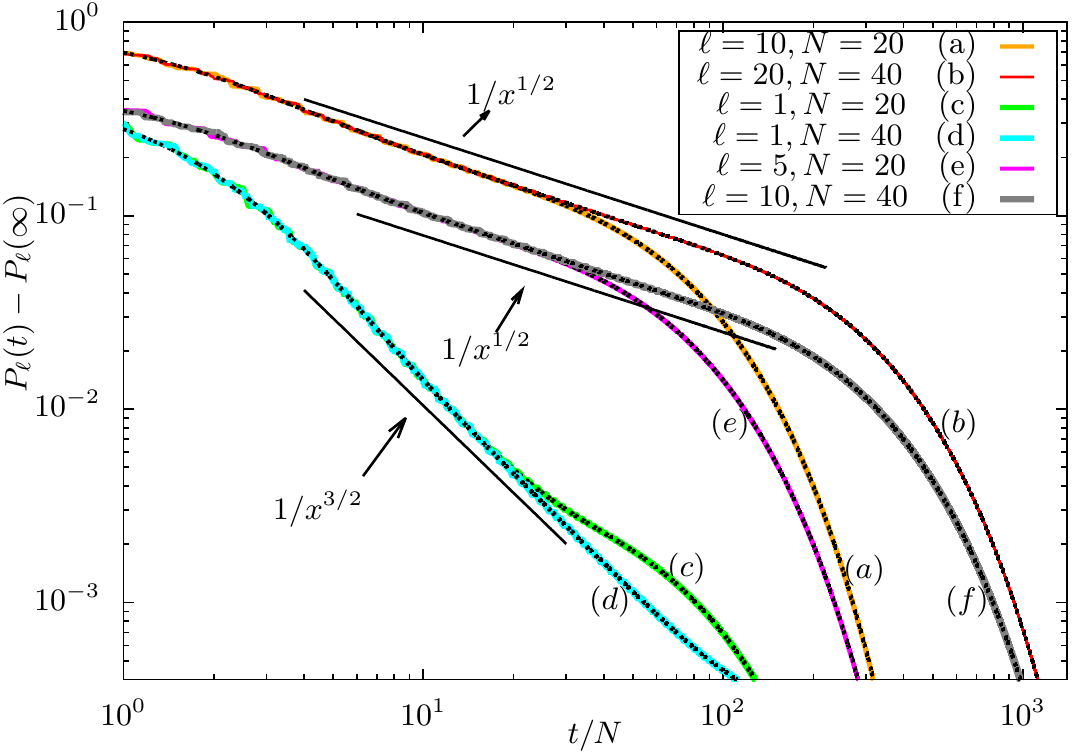}
\caption{(Color online) Periodic Boundary Conditions: Decay of the survival probability 
$P_\ell(t)$ for different initial position eigenstates and different system 
sizes. The upper-most two plots are for $\ell=N/2$ in which case the survival 
probability at infinite times $P_\ell(\infty)$ vanishes, while in all the 
other cases $P_\ell(\infty)=0.5$. The black dotted lines are the predictions 
from perturbation theory. The solid black lines are the predicted power law 
decays for bulk and boundary points. The measurement was done at site $N$, 
and the measurement time interval was taken to be $\tau=0.1$.} 
\label{pbc}
\end{figure}

\begin{figure}
\vspace{-0.7cm}
\hspace{1.5cm}{\includegraphics[clip,width=9cm, angle=0]{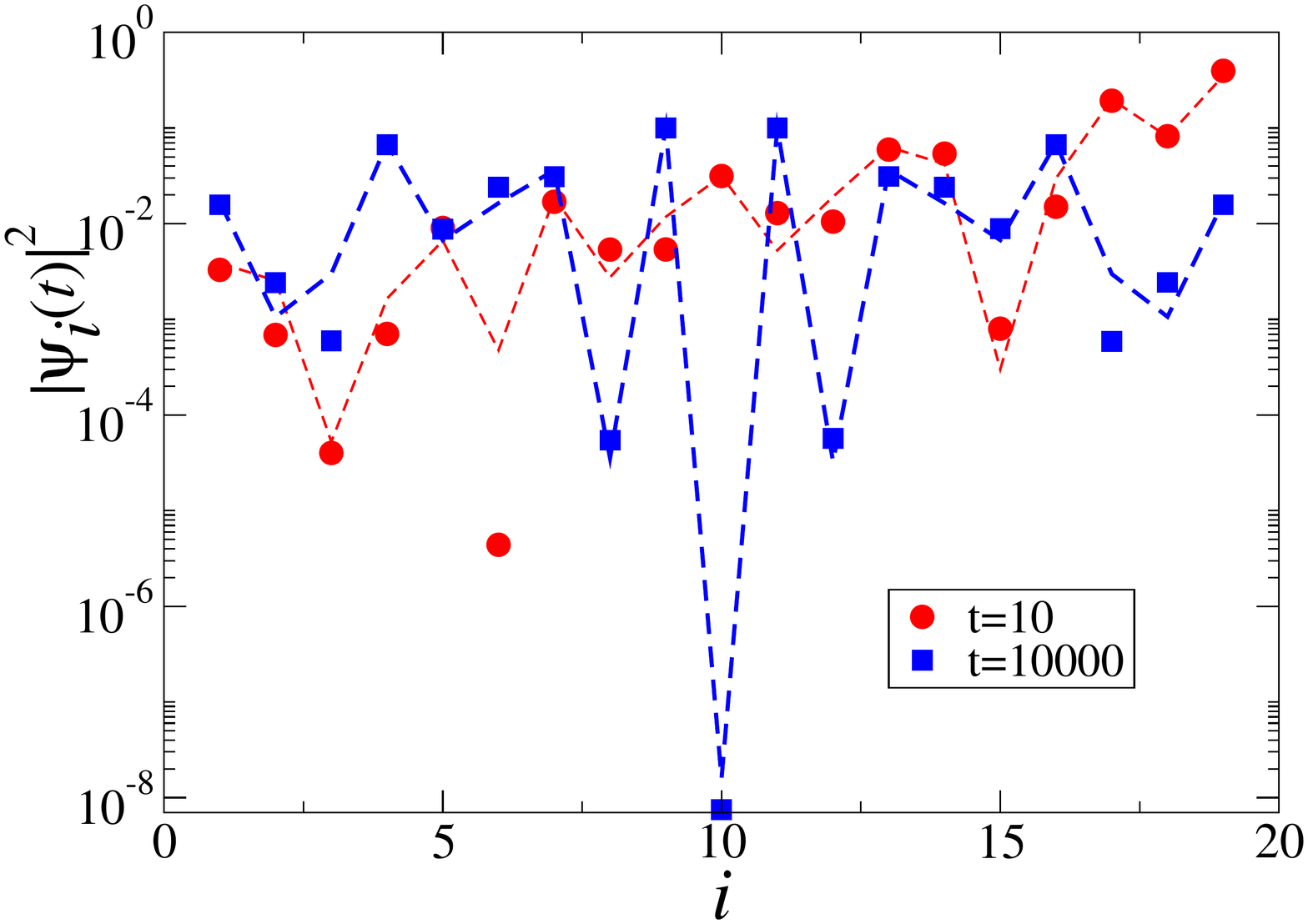}}
{\includegraphics[clip,width=9cm, angle=0]{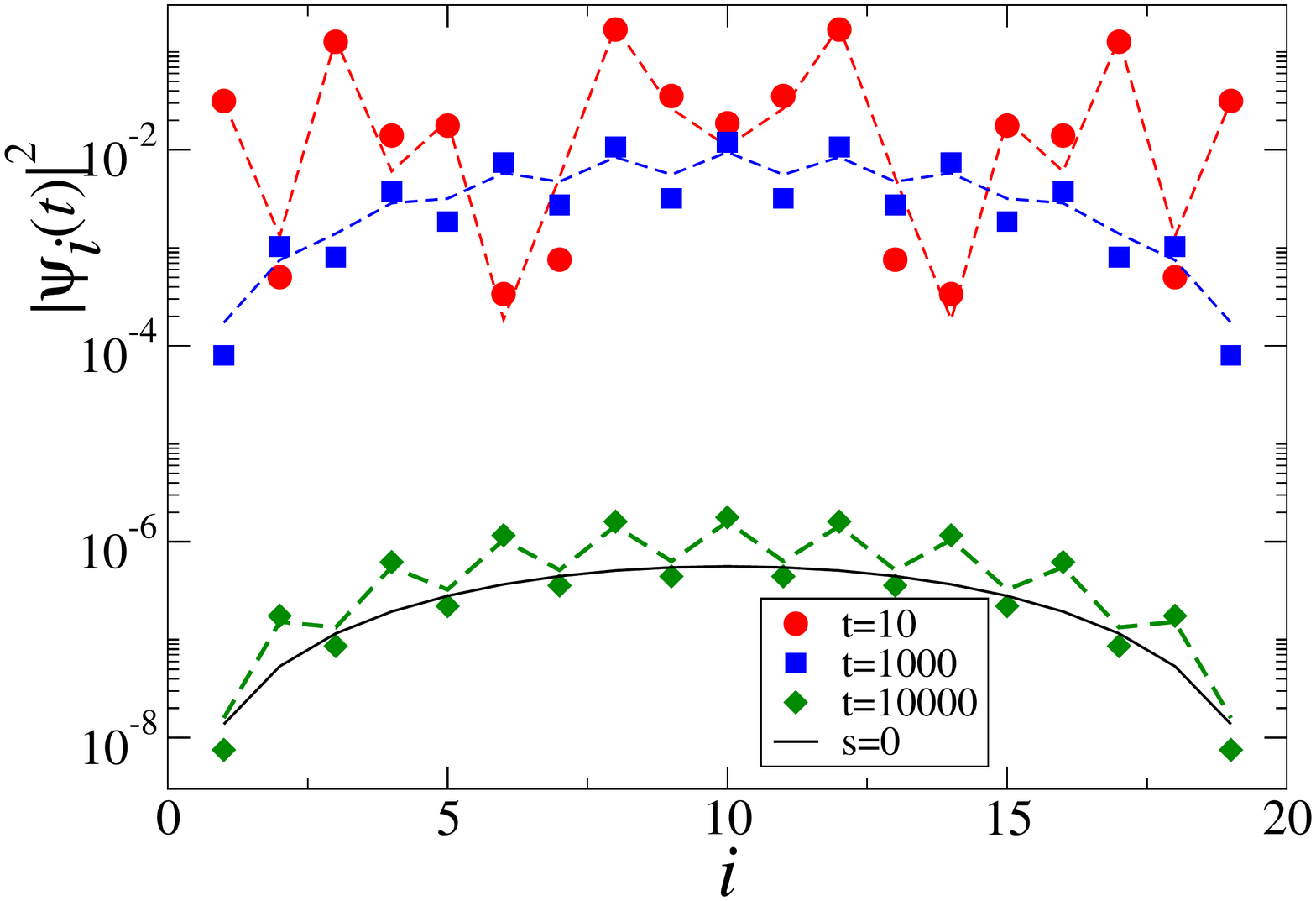}}
\caption{(Color online) Periodic Boundary Conditions: Plot of the probability density 
$|\psi_i(t)|^2$ at different times when the initial state is a position 
eigenstate with (top)$\ell=1$ and (bottom)$\ell=10$. The 
dashed lines are the predictions from the perturbation theory 
[Eq.~(\ref{wfpb})]. The solid line in the bottom panel is the 
plot for the lowest eigenstate. The other parameters were $N=20$, $\tau=0.1$.}
\label{psipbc_l}
\end{figure}

In Fig.~\ref{pbc} we show the comparison between the analytical predictions 
for the survival probability from Eq.~(\ref{Padd1}) and the numerical results 
obtained directly from Eq.~(\ref{PS}); they show reasonably good agreement. 
The form of the wave functions at different times is shown in 
Fig.~\ref{psipbc_l}. We see that as in the case of open boundary conditions,
at large times (${\cal O}(N^3)$) the wave function gets a contribution 
mainly from the surviving eigenstates (for $\ell=1$) or from the lowest 
eigenstate (for the special case $\ell=N/2$). 


\begin{figure}
\includegraphics[clip,width=9cm, angle=0]{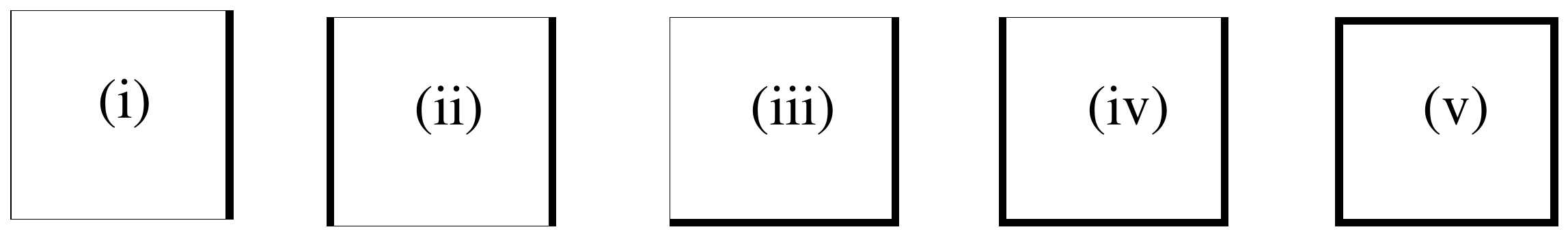}
\caption{Different arrangements of detectors (shown by thick lines) 
for a system on a square lattice. The $(1,1)$ site is at the top left corner, 
and the $x$, $y$ axes are horizontal (to the right) and vertical (downwards).}
\label{schematic2d}
\end{figure}

\subsection{Particle in a two-dimensional box with multiple detection points}
\vspace{-0.5cm}
In this case the Hamiltonian is given by
\begin{align} 
H_{2D}&= - ~\gamma\sum_{l_x=1}^{N-1} \sum_{l_y=1}^{N-1}[ ~|l_x+1, 
l_y \ra \la l_x,l_y | + |l_x,l_y \ra \la l_x+1, l_y |\nonumber \\
& ~~~+ ~|l_x, l_y+1 \ra 
\la l_x,l_y | + |l_x,l_y \ra \la l_x, l_y+1 | ~]~.
\label{2dham}
\end{align}
Again we work with $\gamma=1$. We note that $H_{2D}$ is a sum of 
two commuting Hamiltonians describing jumps along the $x$ and $y$ axes,
\be H_{2D} =H_x + H_y, \ee
where \[H_x = -\sum_{l_x,l_y=1}^{N-1}\left(|l_x+1,l_y\ra \la l_x,l_y| + |l_x,
l_y\ra \la l_x+1,l_y|\right),\] and similarly for $H_y$. Obviously, 
$H_x = H \otimes I$ and $H_y = I \otimes H$, where $H$ is the Hamiltonian for 
an open chain with $N$ sites, and $I$ is an $N\times N$ unit matrix.\\

\no {\em Case (i): Measurement sites placed along one boundary.} \\
We consider measurement sites at $(N,l_y)$ for $l_y=1, 2, \ldots, N$. Then 
$B_{2D} = B\otimes I$, where $B=\sum_{\l=1}^{N-1} |l\ra \la l|$, and
\begin{align} 
{\widetilde{U}}_{2D}&= B_{2D} \exp \left( -iH_{2D}t \right) \nn \\ 
&=\left[ B\exp \left( -iHt \right)\right] \otimes \left[ \exp \left( 
-iHt \right) \right]. 
\label{B2D1}
\end{align}
Hence the behavior of $({\widetilde{U}}_{2D})^n$ is governed by that of 
$[B\exp \left( -iHt \right)]^n$, indicating that the behavior of 
the survival probability of this system is identical with that of an open 
chain (with the same size) with one detector kept at the $N$-th site. 
Moreover, this equivalence (and similar equivalences derived below) is true 
for all values of $\tau$, not necessarily small.\\

\no {\em Case (ii): Measurement sites placed along two opposite boundaries.} 

We consider measurement sites at $(1,l_y)$ and $(N,l_y)$, for $l_y=1, 2, 
\ldots, N$. Then $B_{2D} = B'\otimes I $, where $B'=\sum_{\l=2}^{N-1} |l\ra 
\la l|$, and 
\be {\widetilde{U}}_{2D} = \left[ B'\exp \left( -iHt \right)\right] 
\otimes \left[ \exp \left( -iHt \right)\right]. \label{B2D2}\ee
The behavior of $({\widetilde{U}}_{2D})^n$ is hence governed by that of 
$[B'\exp \left( -iHt \right)]^n$ indicating that the behavior of the 
survival probability of this system is identical with that of an open chain 
with one detector kept at each end (which, in turn, is similar to the survival 
probability of the chain with a detector at the $N$-th site).\\

\no {\em Case (iii): Measurement sites placed along two adjacent boundaries.}

We consider measurement sites at $(l_x,N)$, for $l_x=1, 2, \ldots, N$ and 
$(N,l_y)$, for $l_y=1, 2, \ldots, (N-1)$. Then $B_{2D} = B\otimes B$, and 
\be {\widetilde{U}}_{2D} = \left[ B \exp \left( -iHt \right)\right] 
\otimes \left[ B \exp \left( -iHt \right)\right]. \label{B2D3}\ee
Hence when the initial state is $|\phi_s\ra \otimes |\phi_{s'}\ra $, the 
survival probability becomes
\[P_{s,s'}(t) = e^{-(\alpha_s+\alpha_{s'})t}.\] 
If the particle is initially at $|\ell_x\ra \otimes |\ell_y\ra$, the survival 
probability is
\begin{align}
P_{\ell_x,\ell_y}(t)&=\sum_{s,s'=1}^{N-1}\phi_s^2(\ell_x)\phi_{s'}^2(\ell_y)
P_{s,s'}(t) \label{sp2adjb} \\
&=\f{1}{\sqrt{8\pi x}}[1-e^{-\ell_x^2/2 x}]~\f{1}{\sqrt{8\pi x}}[1-
e^{-\ell_y^2/2 x}], \nn 
\end{align}
with $x=t\tau/N$. Thus the survival probability becomes a product of the 
respective survival probabilities in the two directions, each with one 
detector at the last site.\\ 

\no {\em Case (iv): Measurement sites placed along three boundaries.} 

We consider measurement sites at $(1,l_y)$ and $(N,l_y)$, for $l_y=1,2,\ldots,
N$ and $(l_x,N)$, for $l_x =2, \ldots, (N-1)$. Then $B_{2D} = B'\otimes B$, and
\be {\widetilde{U}}_{2D} = \left[ B' \exp \left( -iHt \right)\right] 
\otimes \left[ B \exp \left( -iHt \right)\right]. \label{B2D4} \ee
When the initial state is $|\phi'_s\ra \otimes |\phi_{s'}\ra $, the survival 
probability is given by
\[P_{s,s'}(t) = e^{-(\alpha'_s+\alpha_{s'})t}. \]
When the particle is initially at $|\ell_x\ra \otimes |\ell_y\ra$, the 
survival probability is
\begin{align}
P_{\ell_x,\ell_y}(t)&=\sum_{s=1}^{N-2}\sum_{s'=1}^{N-1}\phi'^2_s(\ell_x)
\phi_{s'}^2(\ell_y)P_{s,s'}(t) \label{sp3b} \\
&=\f{1}{\sqrt{8\pi x'}}[1-e^{-(\ell_x-1)^2/2 x'}]~\f{1}{\sqrt{8\pi x}}
[1-e^{-\ell_y^2/2 x}],~ \nn 
\end{align}
with $x'=2t\tau/(N-1)$. This is the product of the survival probability for 
a chain with one detector and that for a chain with two detectors at the 
two ends.\\ 

\begin{figure}
\hspace{-.1cm}\includegraphics[clip,width=3.45in, angle=0]{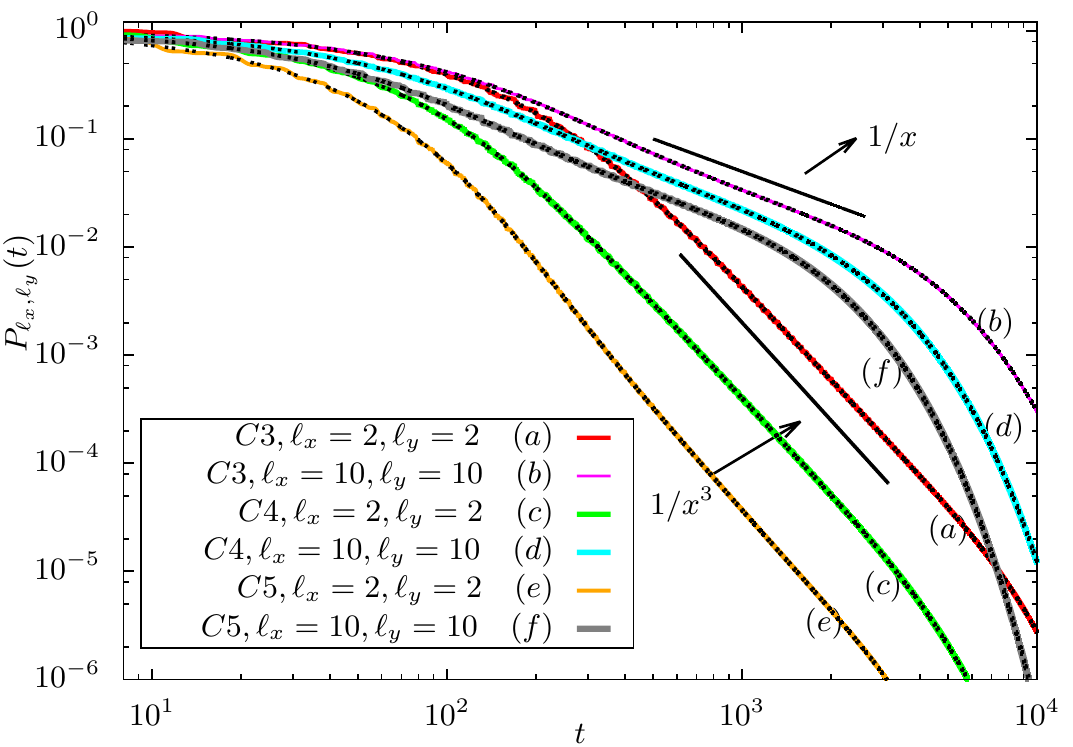}
\caption{(Color online) Decay of the survival probability $P_{\ell_x,\ell_y}(t)$ for 
different initial positions $(\ell_x, \ell_y)$ of a particle moving on a 
square lattice of size $N=20$ with different arrangements of the detectors as 
shown in Fig.~\ref{schematic2d}. C3, C4 and C5 stands for cases (iii), (iv) 
and (v) respectively. The black dotted lines are the predictions from 
perturbation theory. The solid black lines are the predicted power law decays 
for bulk and corner points. (To observe the power law decay for edge points, 
one needs to choose a large size; this is not shown here.) The measurement 
time interval was taken to be $\tau=0.1$.}
\label{ob2dim}
\end{figure}

\no {\em Case (v): Measurement sites placed along all the four boundaries.}

The measurement sites are at $(l_x,1)$ and $(l_x,N)$, for $l_x=1,2,\ldots,N$, 
and at $(1,l_y)$ and 
$(N,l_y)$, for $l_y= 2,3, \ldots,N-1$. Then $B_{2D} = B'\otimes B'$, and 
\be {\widetilde{U}}_{2D} = \left[ B' \exp \left( -iHt \right)\right] 
\otimes \left[ B' \exp \left( -iHt \right)\right]. \label{B2D5}\ee
When the initial state is $|\phi'_s\ra \otimes |\phi'_{s'}\ra $, the survival 
probability is given by
\[P_{s,s'}(t) = e^{-(\alpha'_s+\alpha'_{s'})t}. \]
When the initial state is $|\ell_x\ra \otimes |\ell_y\ra$, the survival 
probability is
\begin{align}
P_{\ell_x,\ell_y}(t)&=\sum_{s,s'=1}^{N-2}\phi'^2_s(\ell_x)\phi'^2_{s'}(\ell_y)
P_{s,s'}(t) \label{sp4b} \\
&= \f{1}{\sqrt{8\pi x'}}[1-e^{-(\ell_x-1)^2/2 x'}]~ \nonumber \\
& ~~~\times \f{1}{\sqrt{8\pi x'}}[1-e^{-(\ell_y-1)^2/2 x'}].~ \nn
\end{align}
This is the product of the survival probabilities of two chains, each with 
two detectors at the two ends. \\

To sum up, if the initial state is a position eigenstate, then for cases (i)
and (ii), the survival probability decays as $t^{-1/2}$ and $t^{-3/2}$, while 
for cases (iii), (iv) and (v), it decays as $t^{-1}$, $t^{-3}$, and $t^{-2}$, 
when the initial position is in the bulk, at the corner (both $x$ and $y$ 
coordinates are near the ends) or at the edge (one of the coordinates is near 
the end, and the other is in the bulk) respectively. In Fig.~\ref{ob2dim} 
we show the comparison between the analytical predictions for the survival 
probability for cases (iii), (iv) and (v) using the expressions in 
Eq.~(\ref{sp2adjb}), (\ref{sp3b}) and (\ref{sp4b}) and the numerical results 
obtained directly from Eq.~(\ref{PS}).

\section{Exactly solvable mean field model}
\label{sec:mf}
We now consider the case when a particle can hop to any of the $N$ sites with 
equal amplitude. Hence the Hamiltonian is
\be H = - \sum_{j,k=1}^N |j \ra \la k |. \ee
For this case, the Hamiltonian matrix $H$ has all elements equal to $-1$, 
and one has $e^{-i{ H}\tau} = I - { H}/c$ where $ c = {N}/{(e^{i\tau N} -1)}$.
The eigenvalues and eigenvectors of $\widetilde{U} =B ~(I-H/c)$ are easily 
found. The eigenvalues are
\bea \lambda_1 = 0,\;\;\lambda_2 = 1+(N-1)/c,\;\;\lambda_s = 
1 ~~{\rm for} ~~s=3,\ldots,N, \nonumber \\
\label{ev} 
\eea
while the corresponding right and left eigenvectors are 
\begin{align} 
|R_1\ra&= \f{1}{1-c-N} (1, 1, \ldots, 1, 1-c-N),~~~\nn \\
|R_2\ra&= \f{1}{N-1} (1, 1, \ldots 1, 0), \nn \\
|R_s\ra&= \f{1}{N-1} (1, \omega_{s}, \omega_{s}^2 \ldots 
\omega_{s}^{N-2}, 0 ) ~~~{\rm for} ~s=3,\ldots,N, \nn \\
\la L_1 |&= (0, 0, \ldots, 0, 1),~~~~\la L_2| = (1, 1, \ldots, 1, 
\f{N-1}{c+N-1}), \nn \\
\la L_s|&= (1, \omega_{s}^{*}, \omega_{s}^{*2}, \ldots, \omega_{s}^{*N-2}, 
0 ) ~~~~~~~{\rm for} ~s=3,\ldots,N, 
\end{align}
and $\omega_s = e^{2\pi i (s-2)/(N-1)}$. Writing $\widetilde{U}=\sum_s 
\lambda_s |R_s \ra \la L_s |$, and taking the initial state to be 
$|\psi(0)\ra =|\ell\ra$ (where $\ell \neq N$), we can use Eqs. (\ref{pdet}) 
and (\ref{PS}) to obtain the first detection probability and the survival 
probability
\begin{align} 
p_n&= \left | \sum_s \la N | U_\tau | R_s \ra ~\lambda_s^{n-1}~ \la 
L_s | \ell \ra \right|^2 \nn \\
&= \left|\f{\lambda_2^{n-1}}{c}\right|^2 = 
\f{x}{N-1}\left[1-x\right]^{n-1}, \label{pn_1} 
\end{align}

\begin{align} 
P_n&= 1- \f{1-(1-x)^n}{N-1}, \;\;\;\nn \\
&{\rm with}~~~ x = \f {2}{N}
\left(1 - \f{1}{N}\right) \left(1- \cos \tau N\right) \label{x_def}~. 
\end{align}

For $\ell=N$, we get $p_1 = 1 - x,~ P_1=x $ and 
\begin{align} 
p_n&= (N-1)^2\left|\f{\lambda_2^{n-2}}{c^2} \right|^2 = x^2 \left[1-x
\right]^{n-2},\;\;\;\; \nn \\
P_n&= x(1-x)^n ~,~~~~~{\rm for}~~ n >1~. \label{pn_2} 
\end{align}
For large $N$, one has $ x \approx 2 (1- \cos \tau N)/N$, and 
\bea p_n & = & \f{2e^{\xi}}{N^2} \left(1- \cos \tau N\right) e^{-n\xi} ~~~
\mbox{for ~$\ell \ne N$} \\
& = & \f{4e^{2\xi}}{N^2} \left(1- \cos \tau N\right) e^{-n\xi} ~~\mbox{for 
$\ell=N$, $n>1$}~ \eea
where $\xi = \f{2}{N} \left(1- \cos \tau N\right)$.

Some properties of the first detection probability can be immediately observed:

\no (1) For $\tau \rightarrow 0$ for a fixed $N$, the probability $p_n$ 
vanishes for all $n$; this is the Zeno effect \cite{misra77,shimizub05,facchi08,wineland90,kofman,kwiat,cirac}.

\no (2) There is a {\em characteristic time} $\tau_0=2\pi/N$ such that 
$p_n(\tau) = p_n(\tau+\tau_0)$. If we choose $\tau = \tau_0 \times 
{\rm integer}$, then $x=0$ according to Eq. (\ref{x_def}). For $\ell \ne N$, 
we then have $p_n = 0$ for all $n$, while for $\ell=N$, $p_1=1$ and 
$p_{n>1}=0$.

\no (3) The quantity $p_n$ decays {\em exponentially with $n$} for all 
values of $N$, large or small (see Eqs. (\ref{pn_1}, \ref{pn_2})). 

\no (4) From Eqs. (\ref{pn_1}, \ref{pn_2}), one can calculate the sum
\be \sum_{n=1}^{\infty} p_n = \left\{ \begin{array}{ll} 
\f{1}{N-1} & ~~\mbox{for ~$\ell \ne N$} \\ 
1 & ~~\mbox{for ~$\ell=N $}. \end{array} \right. \label{sum1} \ee
Thus, if the initial site is different from the detector site then, with a 
finite probability, the particle is never detected. If we compute the wave 
function $|\psi^+_n \ra$ at large times, we find that it converges to 
a ``steady state'' $|\psi_{ss}^+\ra$ with $ \la \ell |\psi^+_{ss}\ra =
(N-2)/(N-1)$ and $ \la j |\psi^+_{ss}\ra =-1/(N-1)$ for all $j \neq \ell,N$. 
This state is invariant under the time evolution, and so we do not see any 
further decay.

\section{Relation to another non-Hermitian Hamiltonian with 
a large imaginary potential}
\label{sec:pert2}

In Sec.~\ref{sec:pert} we showed that the problem of time 
evolution with a Hermitian Hamiltonian, punctuated by measurements at 
intervals of time $\tau$, can be related to continuous time evolution with 
a non-Hermitian Hamiltonian with a small imaginary potential. In this 
section we show that there also exists a connection to another non-Hermitian 
Hamiltonian with a {\emph{large}} imaginary potential. This then relates our 
results to the studies in Refs. \cite{halliwell10,mallick13}.

We will assume that the nearest neighbor hopping amplitude $\gamma$ is the 
same in the two systems; we will not set $\gamma =1$ in this section. 
The formalism in Ref.~\cite{mallick13} contains a non-Hermitian on-site term 
$-i \gamma \Gamma$ at one site ($\Gamma$ being a dimensionless and positive 
real number) which leads to a non-unitary evolution. Assuming that $\Gamma 
\gg 1$ in Ref.~\cite{mallick13} and $\tau \ll 1/\gamma$ in our formalism, we 
will develop second order perturbation theory in these quantities and show 
that the two systems match if $\tau \gamma \Gamma = 2$. In a somewhat more 
general setting of the problem considered in Ref.~\cite{mallick13}, let us 
consider the dynamics of a particle evolving with the following Hamiltonian:
\begin{align}
H_{NH}= H+ \Gamma H' ~~~{\rm{where}}~~ H'=- i\gamma \sum_{\alpha \in D}|\alpha 
\ra \la \alpha |~ \label{nh-ham} 
\end{align}
and $H$ is the Hamiltonian defined in Eq.~(\ref{ham2}).
Note that the non-Hermitian Hamiltonian $H_{NH}$ is for the full system, 
including the sites where measurements are done, unlike the effective 
non-Hermitian Hamiltonian in Eq.~(\ref{effH}). Let us consider the case where 
$\Gamma$ is large and proceed to compute the spectrum of this non-Hermitian 
Hamiltonian using perturbation theory.

Let us write the following expansions for the eigenfunctions and eigenvalues 
of $H_{NH}$:
\begin{align}
| \Psi \ra &= |\Psi_0\ra + \f{1}{\Gamma} |\Psi_1\ra + \f{1}{\Gamma^2} |\Psi_2
\ra +\ldots, \nn \\
E &=\Gamma E_0 + E_1 + \f{1}{\Gamma}E_2 + \ldots \label{pert}~.
\end{align}
This leads to the following equations up to second order in perturbation 
theory:
\begin{align}
H' |\Psi_0\ra &= E_0 |\Psi_0\ra, \nn \\
H' |\Psi_1\ra + H |\Psi_0\ra &= E_0 |\Psi_1\ra +E_1 |\Psi_0\ra, \nn \\
H' |\Psi_2\ra + H |\Psi_1\ra &= E_0 |\Psi_2\ra +E_1 |\Psi_1\ra + E_2 |\Psi_0
\ra. \label{pert2} \end{align}
At $0^{\rm th}$ order, we see that there are $N_s$ degenerate states 
$|l\ra$ with eigenvalues $E_0=0$, and $N_D$ degenerate states $|\alpha\ra$ with 
eigenvalues $E_0=-i \gamma$. We now examine the corrections to the ``system'' 
states at first and second orders in perturbation theory. At first order, the 
eigenstates are given by $|\Psi_0\ra= \sum_l a_l |l\ra,~|\Psi_1\ra=\sum_\alpha 
a_\alpha |\alpha\ra$, and the coefficients and energy corrections are given by
\be \sum_m H_{l,m} a_m = E_1 a_l, ~~~{\rm where} ~~~a_\alpha=\f{1}{i\gamma} \
\sum_m V_{\alpha,m}a_m. \ee
Assuming, for simplicity, that the states $E_1$ are non-degenerate, we see 
that the second order correction is 
\begin{align}
E_2&=\la \Psi_0| H | \Psi_1\ra \nn \\
&= \sum_{l,\alpha} a_l a_\alpha \la l| H |\alpha \ra \nn \\
&= \sum_{l,m} a_l a_m \f{1}{i \gamma} \sum_\alpha V_{l,\alpha} V_{\alpha,m}~.
\end{align}
Thus we see that the energy levels of the ``system'' states $E_1+E_2/\Gamma$ 
are described by the effective Hamiltonian 
\be H_{eff}=H_S-\f{i}{\gamma \Gamma} \sum_{l,m} \sum_\alpha V_{l,\alpha}
V_{\alpha,m} |l\ra \la m|~. \label{effHp} \ee
We note that this is identical to Eq.~(\ref{effH}) with the identification 
\be \f{\tau}{2}=\f{1}{\gamma \Gamma}~.\label{taugamma} \ee
The quantum survival probability for the particle will therefore match in the 
two systems if the conditions $\tau<<1/\gamma$ and $\Gamma>>1$ are satisfied, 
and if Eq.~(\ref{taugamma}) holds.

Earlier in Sec.~\ref{sec:pert} we discussed a mapping of the dynamics of 
the system under repeated measurements to another effective non-Hermitian 
Hamiltonian. Let us clarify the difference between that and the one discussed 
in this section, in the context of the special case of a one-dimensional 
$N$-site chain with open boundary conditions. For this case we have proved 
that the dynamics under repeated measurements at the $N$-th site at time
intervals $\tau$, is identical to the dynamics of both an open chain of
$N$ sites with a strong imaginary potential  $-i2/\tau$ placed at 
the $N$-th site (see Eq.~(\ref{taugamma})) and of an open chain of $(N-1)$ 
sites with a weak imaginary potential $-i\tau/2$ placed at the $(N-1)$-th 
site (see Eq.~(\ref{nh-ham}) and (\ref{H_add})). A corollary of this 
observation is that the dynamics of an open $N$-site chain with a strong 
potential $-iV$ at the $N$-th site is equivalent to that of an open
$(N-1)$-site chain with a weak potential $-i/V$ at the $(N-1)$-th site.

\section{Discussion} 
\label{sec:summary}
We have used a tight-binding model (with a hopping amplitude $\gamma$) for a 
particle on a lattice to study the problem of first detection and survival 
under repeated measurements at a given site or a set of sites. The 
measurements are made at intervals of time $\tau$. We develop a non-unitary 
evolution which describes the probability of first detection of the particle 
at time $t$ and, equivalently, the non-detection or survival of the particle 
up to the time $t$. We summarize our results below. 

Due to the frequent projective measurements made on the system, the wave 
function evolution is non-unitary. We have shown, using a perturbative 
approach, that the dynamics can be described by an effective non-Hermitian 
Hamiltonian, which makes the problem analytically tractable.
For a one-dimensional system with either open or periodic boundary conditions 
and a detector placed at a single site, we derive an analytical expression 
for the survival probability up to a time $t$ using perturbation theory when
$\tau$ is much smaller than the inverse of the band width (proportional to
$\gamma$). If $t$ is
held fixed, we find that the detection probability vanishes in the limit
$\tau \to 0$; this is the quantum Zeno effect. Next, we show that the 
survival probability generally decays as a power of $t$ for a certain range of
values of $t$. The power depends on the initial position of the particle,
namely, whether it is near the detecting point or far away from it; we 
derive an interpolating function which varies from one power law to the other 
as the initial position is changed. (Interestingly, for periodic boundary 
conditions, we find that the survival probability generally approaches a 
non-zero constant as $t \to \infty$). We also find the spatial distribution 
of the particle when it is not detected and show that it approaches a simple 
form as $t \to \infty$. 

We then consider a number of generalizations of the 
model. If we consider an open chain with a number of detection points at one
end, we find that the system effectively behaves like a shorter open chain 
with a single detection point at one end. We study an open chain with a 
detection point at each end and show that the scaling behavior of the survival 
probability is similar to that of an open chain with one detection point. 
We have also studied a particle moving on a two-dimensional square lattice. 
The tight-binding model on this system 
behaves like a product of two decoupled one-dimensional models in the
$x$ and $y$ directions. Hence, the cases in which the detecting sites lie 
along one edge, two edges (which can be adjacent or opposite to each other), 
three edges or all four edges, can be mapped to a product of two open chains 
each of which has a single detecting site at one or both ends. As a result
the survival probability and its scaling with time can be derived easily
in all these cases.
We have then examined a mean field model where the particle can hop between
any pair of sites with the same amplitude. In this system, we find that
the survival probability is generally finite in the large $t$ limit. 

Finally, we have pointed out an interesting connection between our problem and 
another recent work~\cite{mallick13} on the survival probability of a particle 
on a one-dimensional lattice with an imaginary potential at one or more sites.
The latter study uses a non-Hermitian and time-independent Hamiltonian in 
which the potential on a set of sites (which corresponds to the detection 
sites in our formalism) is imaginary and has a value $-i \gamma \Gamma$; hence 
the particle can get absorbed there which is the equivalent of getting 
detected in our language. We show that the two approaches give identical 
results if $\tau$ is much smaller than the inverse band width $1/\gamma$, 
$\Gamma \gg 1$, and $\tau \gamma \Gamma = 2$.

We can consider various extensions of this work for future studies. It may 
be interesting to look at many-body systems and investigate the effect of 
repeated measurements at one point on the particle distribution near that 
point and to see if measurements can give rise to quantum entanglement. It
would also be interesting to look at the effect of measurements of 
observables other than the position, such as the momentum or the spin.

\begin{center}
{\bf Acknowledgments}\\
\end{center}

S. Dhar gratefully acknowledges CSIR, India for providing a research 
fellowship through sanction no. 09/028(0839)/2011-EMR-I. The work of 
S. Dasgupta is supported by UGC-UPE (University of Calcutta). S. Dasgupta is 
also grateful to ICTS, Bangalore for hospitality. AD thanks DST, India for 
support through the Swarnajayanti grant. DS thanks DST, India for 
Project No. SR/S2/JCB-44/2010.

\end{document}